\documentclass[graybox]{svmult}

\usepackage{mathptmx}       
\usepackage{helvet}         
\usepackage{courier}        
\usepackage{type1cm}        
\usepackage{makeidx}         
\usepackage{graphicx}        
\usepackage{multicol}        
\usepackage[bottom]{footmisc}

\makeindex             

\begin{document}

\title*{Progress on the ARIADNE axion experiment}
\author{A.A.  Geraci for the ARIADNE collaboration (A.A. Geraci$^*$, H. Fosbinder-Elkins, C. Lohmeyer, J. Dargert, M. Cunningham, M. Harkness, E. Levenson-Falk, S. Mumford, A. Kapitulnik, A. Arvanitaki, I. Lee, E. Smith, E. Wiesman, J. Shortino, J.C. Long, W.M. Snow, C.-Y. Liu, Y. Shin, Y.Semertzidis, Y.-H. Lee)}
\institute{H. Fosbinder-Elkins, C. Lohmeyer, J. Dargert, M. Cunningham, M. Harkness, A.A. Geraci,  \at Department of Physics, University of Nevada, 1664 N Virgina St., Reno, NV, 89557 \email{$^*$ageraci@unr.edu}
\and E. Levenson-Falk, S. Mumford \at Department of Physics, Stanford University, 382 Via Pueblo, Stanford, CA 94305
\and A.Kapitulnik \at Department of Physics and Applied Physics, Stanford University, 382 Via Pueblo, Stanford, CA 94305
\and A. Arvanitaki \at Perimeter Institute, 31 Caroline St N, Waterloo, ON N2L 2Y5, Canada
\and I. Lee, E. Wiesman, J. Shortino, I. Lee W.M. Snow, J.C. Long, C.-Y. Liu \at Department of Physics, Indiana University, 107 S Indiana Ave, Bloomington, IN 47405 
\and E. Smith \at Los Alamos National Laboratory, Los Alamos, NM 87545
\and Y.Shin, Y. Semertzidis \at IBS Center for Axion and Precision Physics Research, KAIST, 193 Munji-ro, Yuseong-gu, Daejeon 34051, South Korea
\and Y.-H. Lee \at KRISS, 267 Gajeong-ro, Yuseong-gu, Daejeon 34113, Republic of Korea}
%
%
\maketitle

\abstract{The Axion Resonant InterAction Detection Experiment (ARIADNE) is a collaborative effort to search for the QCD axion using techniques based on nuclear magnetic resonance \cite{minaandy}. In the experiment, axions or axion-like particles would mediate short-range spin-dependent interactions between a laser-polarized $^3$He gas and a rotating (unpolarized) tungsten source mass, acting as a tiny, fictitious “magnetic field”.  The experiment has the potential to probe deep within the theoretically interesting regime for the QCD axion in the mass range of 0.1-10 meV, independently of cosmological assumptions.  The experiment relies on a stable rotary mechanism and superconducting magnetic shielding, required to screen the $^3$He sample from ordinary magnetic noise.  Progress on testing the stability of the rotary mechanism is reported, and the design for the superconducting shielding is discussed.}

\section{Introduction}
\label{sec:1}
The axion is a particle postulated to exist since the 1970s to explain the lack of Charge-Parity-violation in the Strong interactions, i.e. the apparent smallness of the angle $\theta_{\rm{QCD}}$ \cite{axion,Moody:1984ba}.   The axion is also a promising Dark Matter candidate \cite{PDG}. In addition, axions or axion-like particles occur quite generically in theories of physics beyond the standard model, and certain compactifications in string theory could give rise to a plenitude of axions with logarithmically distributed masses that give signatures in a wide range of experiments \cite{axiverse}.  Thus the axion belongs to a class of ``economical'' and therefore highly-motivated solutions to some of the greatest puzzles in cosmology and high-energy physics.  The mass of the QCD axion $m_A$ is constrained to lie in a certain range, as shown in Fig. 1. The upper bound comes from astrophysics: white dwarf cooling times and Supernova 1987A data imply $m_A < 6$ meV \cite{PDG}.  The lower bound on the axion mass depends on cosmology \cite{axionCDM,axionCDM2}.  In theories of high-energy scale inflation, the axion mass must not be lighter than about $1$ $\mu$eV to avoid its overproduction and therefore the presence of too much gravitating matter to account for the presently expanding universe.  However, if the energy scale of inflation is low, depending on the initial conditions it is possible that the axion mass could be much lighter than $1$ $\mu$eV, with some lower bounds resulting from black hole superradiance \cite{BH1,BH2}.  ARIADNE \cite{minaandy} will probe QCD axion masses in the higher end of the traditionally allowed “axion window” of $1$ $\mu$eV to $6$ meV, which are not currently accessible by any existing experiment including dark matter ``haloscopes'' such as ADMX \cite{ADMX}.  Thus the experiment fills an important gap in the search for the QCD axion in this unconstrained region of parameter space.

\subsection{Basic principle of experiment}  
The axion can mediate an interaction between fermions (e.g. nucleons) with a potential given by \begin{equation}
 U_{sp}(r)=\frac{\hbar^2 g_s^N g_p^N}{8 \pi m_f}\left( \frac{1}{r \lambda_a}+\frac{1}{r^2}\right) e^{-\frac{r}{\lambda_a}} \left(\hat \sigma \cdot \hat r \right),
 \end{equation}
 where $m_f$ is their mass, $\hat{\sigma}$ is the Pauli spin matrix, $\vec{r}$ is the vector between them, and $\lambda_a = h/m_A c$ is the axion Compton wavelength \cite{Moody:1984ba,minaandy}. For the QCD axion the scalar and dipole coupling constants $g_s^N$ and $g_p^N$ are directly correlated to the axion mass. Since it couples to $\hat{\sigma}$ which is proportional to the magnetic moment of the nucleus, the axion coupling can be treated as a fictitious “magnetic field” $B_{\rm{eff}}$.  This fictitious field is used to resonantly drive spin precession in a sample of laser polarized cold $^3$He gas.  This is accomplished by spinning an unpolarized tungsten mass sprocket near the $^3$He vessel. As the teeth of the sprocket pass by the sample at the nuclear Larmor precession frequency, the magnetization in the longitudinally polarized He gas begins to precess about the axis of an applied field. This precessing transverse magnetization is detected with a superconducting quantum interference device (SQUID). The $^3$He sample acts as an amplifier to transduce the small, time-varying, fictitious magnetic field into a larger real magnetic field detectable by the SQUID. Superconducting shielding screens the sample from ordinary magnetic noise which would otherwise exceed the axion signal, while not attenuating $B_{\rm{eff}}$ \cite{minaandy}. The ultimate sensitivity limit is set by spin projection noise in the sample itself which scales with the inverse square root of its volume, density, and $T_2$, the transverse spin decoherence time \cite{minaandy}. 
 
 The experiment can sense all axion masses in its sensitivity band simultaneously, while haloscope experiments must scan over the allowed axion oscillation frequencies (masses) by tuning a cavity \cite{ADMX,x3,orpheus,LCcircuits} or magnetic field \cite{casper}. In contrast to other lab-generated spin-dependent fifth-force experiments using magnetometry \cite{Vasilakis:2008yn,gsgpgermany,snow,snow2,lamoreauxhunter}, the resonant enhancement technique affords orders of magnitude improvement in sensitivity, sufficient to detect the QCD axion (Fig. 1).

\begin{figure}[!t]
\begin{center}
\includegraphics[width=1.05\columnwidth]{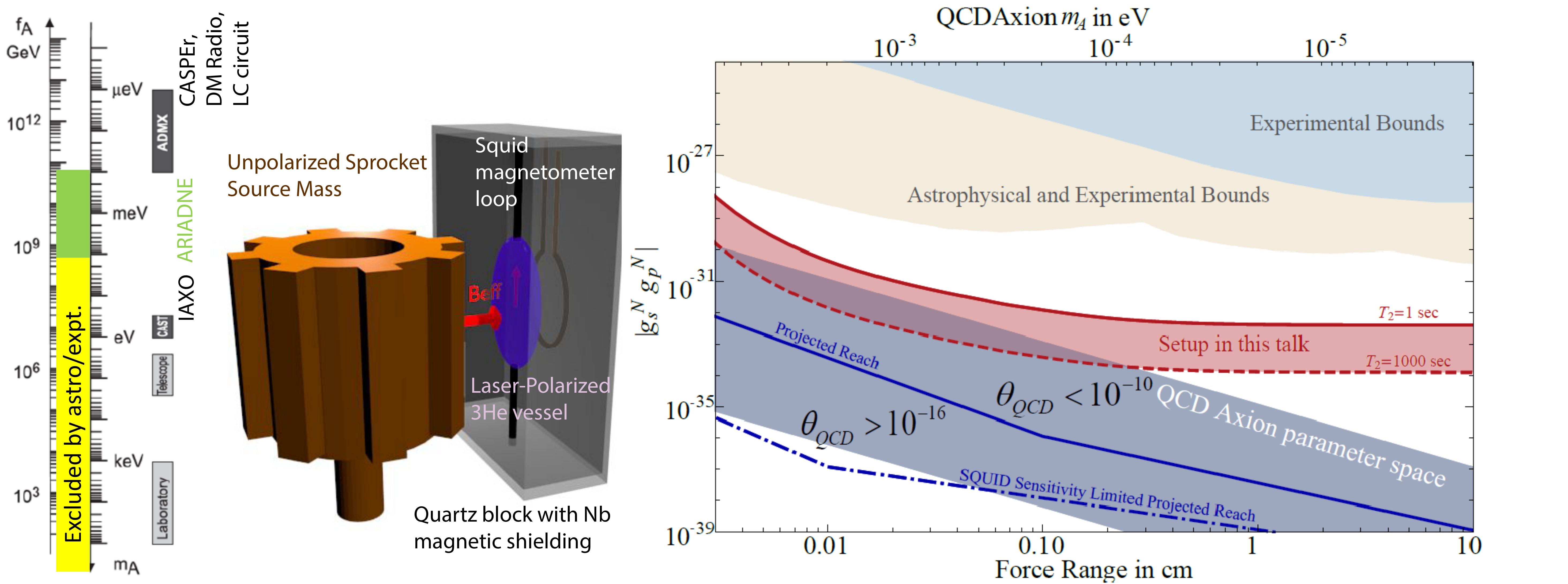}
\caption{(left) Constraints and experiments searching for the QCD axion, adapted from Ref. \cite{PDG}. (middle) Setup: a sprocket-shaped source mass is rotated so its ``teeth'' pass near an NMR sample at its resonant frequency. (right) Estimated reach for monopole-dipole axion mediated interactions. The band bounded by the red (dark) solid line and dashed line denotes the limit set by transverse magnetization noise, depending on achieved $T_2$ for an integration time of $10^6$ s. The solid ``projected reach'' curve represents the sensitivity of a future, upgraded apparatus \cite{minaandy}. Current constraints and expectations for the QCD axion also are shown \cite{Raffelt:2012sp,Vasilakis:2008yn,gsgpgermany,snow,snow2,lamoreauxhunter}.}
\end{center}
\end{figure}

\section{Experimental Design}
\label{sec:2}

\emph{Detectors.} For detecting the effective magnetic field produced by the axion $B_{\rm{eff}}$, three fused quartz vessels containing laser-polarized $^3$He will serve as resonant magnetic field sensors. Three such sensors will be used to cancel common-mode noise by correlating their signals, according to the phase of the rotation of the sprocket source mass. Each sample chamber has an independent bias field control to maintain resonance between the spinning mass and the $^3$He, during any gradual demagnetization due to the finite $T_1$ time. A SQUID pickup coil will sense the magnetization in each of the samples. The inside of the quartz containers will be polished to a spheroidal shape with principal axes $10$ mm $\times$ $3$ mm $\times$ $150$ $\mu$m. The spheroidal shape will allow the magnetization to remain relatively constant throughout the sample volume, since the magnetization direction will remain reasonably well-aligned with the principal axes \cite{ellipsoid}. The cavity is fabricated by fusing together two pieces of quartz containing hemi-spheroidal cavities \cite{precisionglass}. The vessel wall thickness is limited to $75$ $\mu$m on one side to allow close proximity to the source mass \cite{precisionglass}.

\begin{figure}[!t]
\begin{center}
\includegraphics[width=1.0\columnwidth]{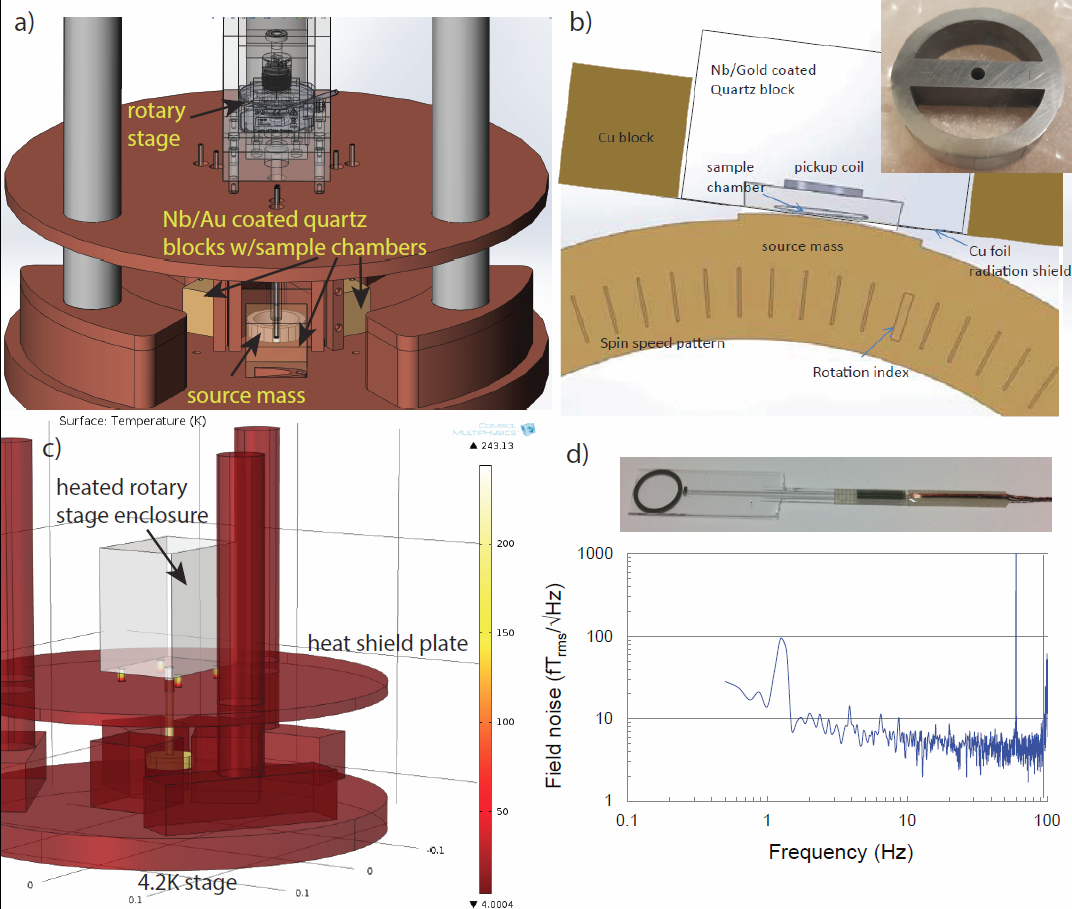}
\caption{(a) Schematic of cryostat bottom plate. The three sample regions are enclosed in quartz blocks which are coated with Niobium followed by Ti/Cu/Au to lower their emissivity.  The blocks are attached to a Cu cold stage (4.2K).
A heated Au-coated enclosure houses the ultrasonic piezoelectric high-speed rotation stage, and heat shields isolate this region from the quartz block. A source mass sprocket with $11$ sections is rotated around at a frequency $\omega_{\rm{rot}}$, which results in a resonance between the frequency $\omega = 11 \omega_{\rm{rot}}$ at which the segments pass near the
sample and the NMR frequency $2 \vec \mu_{N} \cdot \vec B_{ext} /\hbar $, set by the magnetic field. (b) Cross sectional view from top of region near mass and detector. (inset) prototype W rotor.
(c) Thermal model of cryostat using COMSOL. Results indicate an expected He boil-off of $<1.5$ L/hr for maintaining the motor at $-30^{\circ}$ C. (d) (top) prototype SQUID magnetometer fabricated on quartz substrate. (bottom) Magnetic field sensitivity of SQUID magnetometer on Si substrate fabricated at KRISS. \label{probefig}}
\end{center}
\end{figure}

\emph{Source mass.} The rotating source mass consists of a ``sprocket'' of height $1$ cm, inner diameter $3.4$ cm, and outer diameter $3.8$ cm, divided into $22$ sections of length $5.4$ mm. The section radii are modulated by approximately $200$ $\mu$m in order to generate a time-varying potential at frequency $\omega = 11 \omega_{\rm{rot}}$, due to the difference in the axion potential as each section passes by the sensor. The factor of $11$ difference between $\omega_{\rm{rot}}$ and $\omega$ decouples mechanical vibration from the signal of interest.
The sprocket will be driven by a ceramic shaft and precision ceramic bearings. The wobble of the sprocket will be measured in-situ using fiber coupled laser interferometers and counterweights
will be applied as necessary to maintain the wobble below 0.003 cm at the outer radius. An image of a prototype tungsten sprocket appears in Fig. \ref{probefig}. Magnetic impurities in the source mass are estimated to be below the $0.4$ppm level based on measurements of similar machined material using a commercial SQUID magnetometer system \cite{MPMS}.

\emph{Rotation stage.} The cylinder will be rotated by an in-vacuum piezoelectric transducer \cite{fukoku} or direct-drive stage \cite{direct}. Direct drive stages \cite{direct} offer faster rotation rates (up to $25$ Hz) with the caveat that local magnetic fields are larger. These can be attenuated with additional $\mu$-metal shielding (which lies well outside of the superconducting coating on the quartz sample vessels) The rotational mechanism will need to be maintained at a higher temperature ($-30$ C) than the surrounding components. This will be achieved with heaters and heat-shielding, as schematically indicated in Fig. 2. This design results in an expected heat load of approximately 1 W and by suitable thermal isolation we expect the excess evaporation rate of helium due to this additional heat load to be $<1.5$ L/hr.

\begin{figure}[!t]
\begin{center}
\includegraphics[width=1.0\columnwidth]{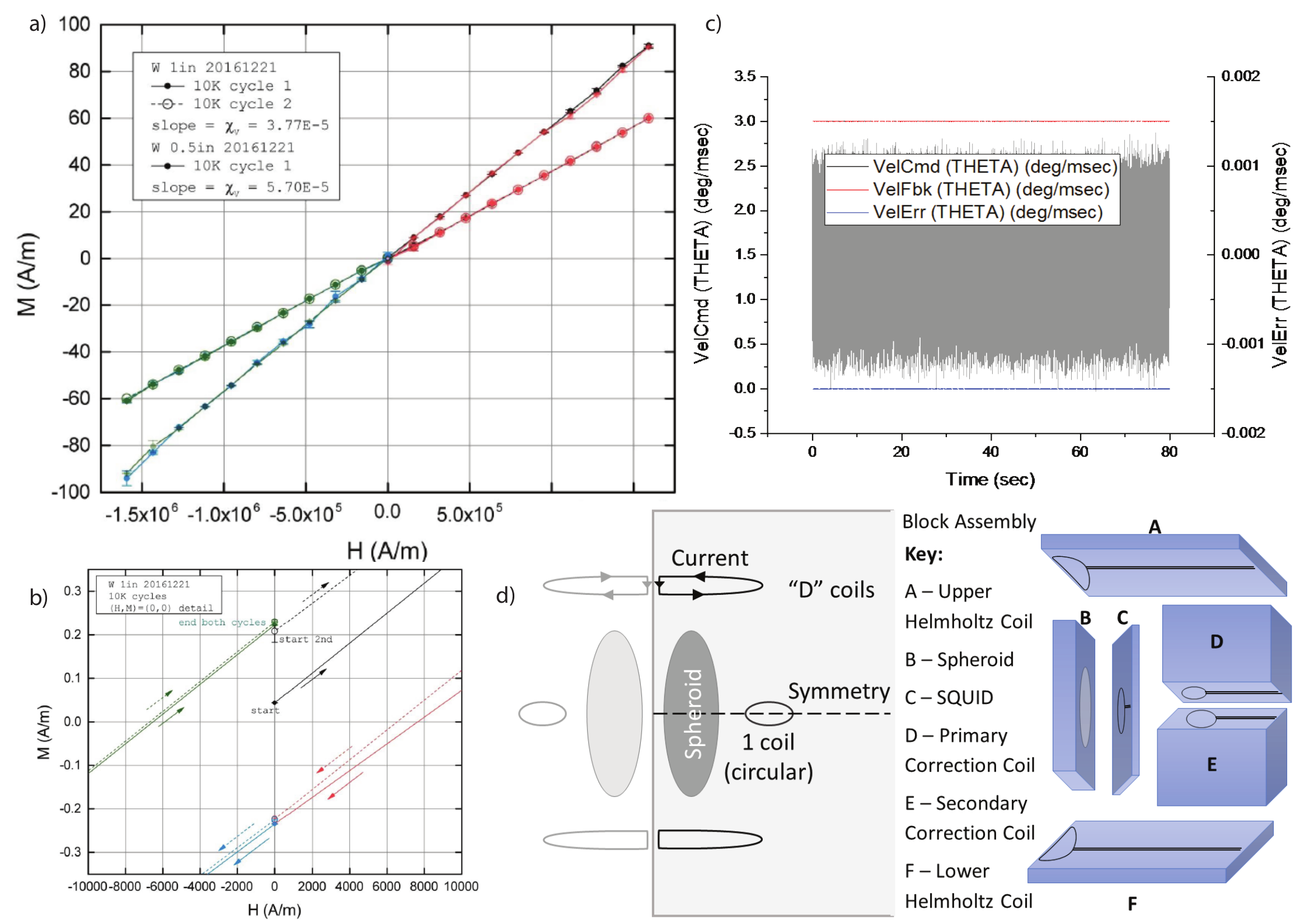}
\caption{
(a) Magnetization test for impurities in W material used to fabricate rotor using MPMS SQUID measurement system \cite{MPMS}. (b) zoom-in of near zero field region. (c) Spin-speed stability of Aerotech ADRS-100 direct drive stage unloaded and in-air as measured via an optical encoder. (d) (left) A symmetrically placed coil can be used to mitigate gradients across the cell produced by the Meissner ``image'' of the spheroid. Two ``D''-shaped coils can be used to approximate a Helmholtz coil near a superconducting boundary. (right) Schematic of quartz block construction prior to Nb coating, showing $^3$He sample block as well as patterned coils on additional quartz sections. \label{rotor}}
\end{center}
\end{figure}

The rotational speed must be kept constant, so that resonance between the Larmor frequency and rotational frequency can be maintained. The spin speed can be measured using an optical encoder. An index mark will allow the determination of the phase of the cylinder rotation, for correlation with the spin precession in the sample. Preliminary tests of the rotation speed stability of the vacuum-prepared unloaded direct-drive stage in-air indicate constant rotation speed at the part in $10^4$ level at the frequency of interest with rms variation at $\sim 1$ part in $3000$, as shown in Fig. \ref{rotor}. This allows the sample to remain on resonance and utilize $T_2 > 100$ s. 

\emph{Cryostat.} The samples will be housed in a liquid Helium cryostat (Fig. 2a). The separation between the rotating mass and quartz cell will be designed to be $75$ $\mu$m at cryogenic temperature. A thermal model of the heat load appears in Fig 2c. A stretched copper radiation shield foil of dimensions $25$ $\mu$m $\times$ $1$ cm $\times$ $1$ cm will be inserted between the rotating mass and the quartz block. The foil will be supported by larger copper blocks affixed to the cold plate of the cryostat. 

For $\omega_{\rm{rot}}/2\pi = 10$ Hz, the net $B_{\rm{ext}}$ needed at the sample is of order 30 mG. $B_{\rm{ext}}$ is the sum of the internal magnetic field of the sample, which is roughly $0.2$~Gauss for $2\times 10^{21}$ cm$^{-3}$ density of $^3$He, and a field generated by superconducting coils. In such a field, the SQUID can operate near its optimal sensitivity of 1.5~fT/$\sqrt{Hz}$.  We expect the current in the coils needs to be maintained constant at low frequencies to within $\sim 10$ ${\rm{ppm}} (1000 {\rm{s}}/T_2)$. 

\emph{Magnetic shielding.} The fictitious field from the QCD axion coupling is at or below the $10^{-19}$ T level, necessitating the shielding of magnetic backgrounds. The quartz sample container will be affixed to a larger quartz block, which will be sputter coated with a $1.5$ $\mu$m layer of Niobium and 200 nm of Ti/Cu. The Cu will then be electroplated with $1$ $\mu$m of Gold as a blackbody reflector. The use of superconducting shielding (as opposed to e.g. $\mu$-metal shielding) is essential to mitigate magnetic field noise from thermal currents (i.e. Johnson noise) in ordinary conducting materials. The shield also attenuates magnetic noise due to thermal currents in the tungsten mass \cite{varpula}, which we estimate at $10^{-12}$ T$/\sqrt{\rm{Hz}}$, and screens the Barnett effect \cite{barnett} along with the effects of magnetic impurities and the nonzero magnetic susceptibility of the source mass rotor. A magnetic shielding factor $f = 10^8$ at the nuclear Larmor frequency $\sim 100$ Hz would allow full design sensitivity to be attained for a $T_2$ of $1000$ s. A list of the these requirements is shown in Table I. In addition, a triple-layer $\mu$-metal shield will enclose the cryostat while the superconducting shields are cooled through the superconducting phase transition.  The total DC magnetic field should be kept below $10^{-7}$ T during this process to avoid ``freezing-in'' flux.

\emph{$^3$He delivery.} The hyperpolarized $^3$He gas is prepared by metastability exchange optical pumping (MEOP). MEOP is especially well-suited to the needs of this experiment: it can polarize $^{3}$He at total pressures of a few mbar in an arbitrary mixture of $^{3}$He and $^{4}$He. The MEOP polarized $^{3}$He compression system at Indiana can deliver polarized $^{3}$He gas at pressures from 1 mbar to 1 bar at room temperature \cite{snowrsi} and can therefore be used to conduct higher temperature tests in the same pressure regime that the 4K cryogenic cell will operate in.  A manifold is under development to deliver the gas from the polarization region into the sample area, and then to recirculate it for subsequent experimental cycles. The low spin relaxation valve technology needed to do this both at room temperature and at low temperature has been developed for neutron spin filters at neutron scattering facilities~\cite{tong} and for the neutron EDM project~\cite{beck}.

\section{Expected sources of systematic errors and noise}
\label{sec:3}

\begin{table}[!t]
\small
\begin{center}

   \begin{tabular}{@{}ccc@{}}
  \hline
  \hline
  Systematic Effect/Noise source & Background Level & Notes \\
  \hline
Magnetic gradients& $3\times 10^{-6}$ T/m &  Limits $T_2$ to $\sim 100$ s\\
&  &  Possible to improve w/SC coils\\
Vibration of mass& $10^{-22}$ T & For $10$ $\mu$m mass wobble at $\omega_{\rm{rot}}$\\
External vibrations& $5 \times 10^{-20}$ T$/\sqrt{{\rm{Hz}}}$ & For $1$ $\mu$m sample vibration ($100$ Hz)  \\
Patch Effect& $10^{-21} (\frac{V_{\rm{patch}}}{0.1 {\rm{V}}})^2$ T & Can reduce with $V$ applied to Cu foil \\
Flux noise in squid loop & $2 \times 10^{-20}$ T$/\sqrt{{\rm{Hz}}}$ & Assuming $1 \mu \Phi_0 /\sqrt{{\rm{Hz}}}$ \\
Trapped flux noise in shield& $7 \times10^{-20} \frac{\rm{T}}{\sqrt{\rm{Hz}}}$ & Assuming $10$ cm$^{-2}$ flux density  \\
Johnson noise& $10^{-20} (\frac{10^8}{f}) {\rm{T}}/\sqrt{{\rm{Hz}}} $ & $f$ is SC shield factor (100 Hz) \\
Barnett Effect& $10^{-22} (\frac{10^8}{f}) $ T & Can be used for calibration above 10 K \\
Magnetic Impurities in Mass & $10^{-25}-10^{-17} (\frac{\eta}{1 {\rm{ppm}}}) (\frac{10^8}{f}) $ T & $\eta$ is impurity fraction \\
Mass Magnetic Susceptibility & $10^{-22} (\frac{10^8}{f}) $ T & Assuming background field is $10^{-10}$ T \\
 &  & Background field can be larger if $f>10^8$ \\

  \hline
  \hline
  \end{tabular}

\caption{\label{table1} Table of estimated systematic error and noise sources, as discussed in the text. The projected sensitivity of the device is $3 \times 10^{-19} (\frac{1000 {\rm{s}}}{T_2})^{1/2}$ T/$\sqrt{\rm{Hz}}$ }
\end{center}
\end{table}

{\emph{Magnetic Gradients and Frequency Shifts}}. In order for the full sample to remain on resonance, gradients across the sample need to be controlled at the challenging level of $\sim 10^{-11} \left( \frac{1000~\mbox{s}} {T_2}\right)$ T. A spheriodal chamber of uniformly magnetized gas results in a constant magnetic field in the interior.  Thus the spheroidal shape of the sample suppresses magnetic gradients due to the magnetized gas itself. However, gradients and frequency shifts can be produced due to image currents arising from the Meissner effect in the superconducting shield.  

A superconducting coil setup can also partially cancel the gradient,
allowing extension of $T_2$ up to $100$ s for a $99\%$ compensation \cite{RSIinprep}. It is useful to conceptually visualize the Meissner effect as producing an spheroid of ``image dipoles'' on the other side of the superconducting boundary.
SC coils are included in the design of the quartz block, to allow either direct cancellation (where the coil field opposes the image spheroid field at the sample)
 or ``filling in'' of the field from the image dipole (where the coil field adds to the field from the image dipole to remove gradients). 
 We calculate that gradients can be suppressed in the central $80\%$ region
of the sample by approximately $\sim 3 \times$ using the field cancellation method, or by $99 \times$ using the gradient cancellation method, where the
field from the image dipole is ``filled in'' with the coil field \cite{RSIinprep}. Two coils are included in the quartz
block design so both approaches can be implemented in the experiment (see Fig. \ref{rotor}).

\emph{Bias field control}. In order to set the nuclear Larmor precession frequency at the $^3$He sample, it is necessary to apply a constant bias field to the spheroid. The ordinary approach involving a Helmholtz coil will not work due to the close proximity of the superconducting boundary.  However we can exploit the Meissner effect to our advantage by using ``D''-shaped coils to create a Helmholtz like field, as shown in Fig. \ref{rotor}.  Here the field from the Meissner image of the coil adds to that from the original coil to produce a nearly constant field at the location of the spheroid \cite{RSIinprep}.

{\emph{Acoustic vibrations}}. Acoustic vibrations can cause magnetic field variations due to the image magnetization arising from the Meissner effect in the superconducting shields. Although $\omega_{\rm{rot}} << \omega$, vibrations can in principle be transmitted at $\omega$ from nonlinearities.  For a $10$ $\mu$m wobble in the cylinder at $\omega_{\rm{rot}}/2\pi = 10$ Hz, and $1$ percent of this at 100 Hz, we estimate a $\delta_x \sim 2$ nm vibrational amplitude of the sample chamber. 
Assuming $\delta_x =2$ nm, we find that the relative motion between the sample chamber and the shield coating on the outside surface of the quartz block from elastic deformations is $10^{-17}$ m.  With a gradient of $10^{-5}$ T/m, this corresponds to a field background of $\sim 10^{-22}$ T.  While this background could principle be coherent, as it is associated with the rotation of the cylinder, at this level it will not dominate over any detectable axion signal for $10^6$ seconds of averaging time. 

Background vibration also should remain below $10$ $\mu$m amplitude at $100$ Hz, since this would produce magnetic field noise of $5 \times 10^{-19}$ T$/\sqrt{\rm{Hz}}$ at the resonant frequency, which can in principle begin to limit the sensitivity. 

{\emph{Trapped Flux}}.
The thermal noise from a trapped flux at distance $r$ from the sample will produce a field noise of $7 \times10^{-20} \frac{T}{\sqrt{Hz}} \left(\frac{200~\mu m}{r}\right)^3$ \cite{blas-pinningforce, simanek-vortexmass}. This estimate is indicative and the actual noise from trapped flux will depend on the construction of the shield. In principle the experiment can tolerate a ``frozen-in'' DC field as large as $10^{-7}$ T with little deleterious effects, however the exact magnetic noise from moving flux at $100$ Hz will need to be experimentally characterized for the particular shield. 
If necessary, additional $\mu$-metal or cryoperm \cite{cryoperm} shielding layers may be included inside and outside of the cryostat. 

{\emph{Patch Potentials}}.
The sputtered metal films which coat the block will generally be polycrystalline, and thus will exhibit local regions of varying work function and hence local contact potential differences \cite{patch}. 
Fluctuating electric patch potentials can drive an oscillating time varying force on the copper heat shield membrane between the quartz sample container and the rotating mass. Assuming a $100$ mV periodic signal, the force on the copper membrane will be $\sim 1.4 \times 10^{-8}$ N, resulting in $\sim 5$ pm of vibration. If this copper foil has an additional $100$ mV potential difference with respect to the gold/Nb coated quartz, this can drive the quartz sample block with a force of $6 \times 10^{-15}$ N.   This in turn can vibrate the sample container's thin quartz wall with a very small amplitude of $0.5$ fm. With a gradient of $10^{-5}$ T/m, this can cause a magnetic field background on resonance of $5 \times 10^{-21}$ T. A voltage can be applied to the copper foil to minimize the DC component of such coupling, although smaller local variations can remain \cite{patch2}. 
By slightly increasing the thickness of either the copper foil or the quartz container wall, we can increase their stiffness and thus further suppress the effect if necessary.

\section{Discussion and Outlook}
ARIADNE is a new approach towards discovering the QCD axion or axion-like particles in a mass range which is larger than that currently being probed in Dark Matter axion experiments. It is similar in style to the light shining through walls experiments such as ALPS or ALPS-II \cite{ALPS} in that (virtual) axions are produced in the lab, except it probes the axion scalar and dipole coupling to nuclei rather than to photons. Simulations and experimental tests conducted thus far indicate that several of the key requirements of the experiment are within reach, including the rotary-stage speed stability, thermal management for the cryostat, and magnetic gradient compensation strategy. Further experimental tests are underway with regard to the testing thin-film superconducting magnetic shielding, the $^3$He polarization and delivery system, metrology of the masses and sample enclosures, SQUID magnetometer system, and rotational mechanism. If successful, the experiment has the potential solve the strong-CP problem and identify a particle which makes up part (or all) of the Dark Matter in the universe.

\begin{acknowledgement}
We thank S. Koyu and H. Mason for computer modeling simulations at the early stages of this work. We acknowledge support from the U.S. National Science Foundation, grant numbers NSF-PHY 1509805, NSF-PHY 1510484, NSF-PHY 1509176. I. Lee, C-Y. Liu, J. C. Long, J. Shortino, W. M. Snow, and E. Weisman acknowledge support from the Indiana University Center for Spacetime Symmetries.                                                              
 
\end{acknowledgement}
%
%
%


%
%

\begin{thebibliography}{99}

\bibitem{minaandy} A. Arvanitaki and A. Geraci, Phys. Rev. Lett. 113, 161801 (2014).





\bibitem{axion}
  R.~D.~Peccei and H.~R.~Quinn,
  Phys.\ Rev.\ Lett.\  {\bf 38}, 1440 (1977);
  S.~Weinberg,
  Phys.\ Rev.\ Lett.\  {\bf 40}, 223 (1978);
  F.~Wilczek,
  Phys.\ Rev.\ Lett.\  {\bf 40}, 279 (1978).

\bibitem{Moody:1984ba}
  J.~E.~Moody and F.~Wilczek,
  Phys.\ Rev.\ D {\bf 30}, 130 (1984).


  P.~Svrcek and E.~Witten,
  JHEP {\bf 0606}, 051 (2006)
  [hep-th/0605206].

\bibitem{axiverse}
A. Arvanitaki, S. Dimopoulos, S. Dubovsky, N. Kaloper, J. March-Russell, Phys. Rev. D81 ,123530 (2010).

\bibitem{PDG}
  J.~Beringer {\it et al.}  [Particle Data Group Collaboration],
  ``Review of Particle Physics (RPP),''
  Phys.\ Rev.\ D {\bf 86}, 010001 (2012).

  \bibitem{axionCDM} L. Visinelli and P. Gondolo,  Phys. Rev. Lett {\bf{113}} 011802 (2014).

 \bibitem{axionCDM2} David J. E. Marsh, Daniel Grin, Renée Hlozek, and Pedro G. Ferreira, Phys. Rev. Lett. 113, 011801 (2014).

\bibitem{BH1}
A. Arvanitaki, S. Dubovsky, Phys.Rev. D83, 044026 (2011).
\bibitem{BH2}
A. Arvanitaki, M. Baryakhtar, X. Huang, Phys.Rev. D91, 084011 (2015). 


\bibitem{ADMX} S. J. Asztalos, G. Carosi, C. Hagmann, D. Kinion, K. van Bibber, M. Hotz, L. J Rosenberg, G. Rybka, J. Hoskins, J. Hwang, P. Sikivie, D. B. Tanner, R. Bradley, and J. Clarke, Phys. Rev. Lett. {\bf{104}} 041301 (2010).





\bibitem{Raffelt:2012sp}
  G.~Raffelt,
  Phys.\ Rev.\ D {\bf 86}, 015001 (2012).

\bibitem{Vasilakis:2008yn}
  G.~Vasilakis, J.~M.~Brown, T.~W.~Kornack and M.~V.~Romalis,
Phys.\ Rev.\ Lett.\  {\bf 103}, 261801 (2009).

\bibitem{gsgpgermany}
K. Tullney, F. Allmendinger, M. Burghoff, W. Heil, S. Karpuk, W. Kilian, S. Knappe-Grüneberg, W. Müller, U. Schmidt, A. Schnabel, F. Seifert, Yu. Sobolev, and L. Trahms, 
 Phys.\ Rev.\ Lett.\  {\bf 111}, 100801 (2013)

 \bibitem{snow}
 P.-H. Chu, A. Dennis, C. B. Fu, H. Gao, R. Khatiwada, G. Laskaris, K. Li, E. Smith, W. M. Snow, H. Yan, and W. Zheng, Phys. Rev. {\bf{D 87}}, 011105(R) (2013).

 \bibitem{snow2} M. Bulatowicz, R. Griffith, M. Larsen, J. Mirijanian, C. B. Fu, E. Smith, W. M. Snow, H. Yan, and T. G. Walker,  Phys. Rev. Lett. 111, 102001 (2013).

\bibitem{lamoreauxhunter} A. N. Youdin, D. Krause, Jr., K. Jagannathan, L. R. Hunter, and S. K. Lamoreaux,  Phys. Rev. Lett. 77, 2170 (1996).







\bibitem{x3} Karl van Bibber and Gianpaolo Carosi, 8th Patras Workshop on Axions, WIMPs, and WISPs, Chicago, IL, July 18-22, arxiv: 1304.7803 (2012).

\bibitem{orpheus} Gray Rybka, Andrew Wagner, Kunal Patel, Robert Percival, Katleiah Ramos, and Aryeh Brill, Phys. Rev. D 91, 011701(R) (2015).
\bibitem{LCcircuits} B. Cabrera and S. Thomas, Workshop Axions 2010, U. Florida, 2010.

  \bibitem{casper}
  D.~Budker, P.~W.~Graham, M.~Ledbetter, S.~Rajendran and A.~Sushkov,
  Phys. Rev. X 4, 021030 (2014).
































\bibitem{precisionglass} private communication, Ron Bihler, Precision Glass Blowing, www.precisionglassblowing.com

\bibitem{ellipsoid} M. Tejedor, H. Rubio, L. Elbaile, and R. Iglesias, IEEE Trans. Magnetics, 31, 830 (1995).


\bibitem{varpula} T. Varpula and T. Poutanen, J. Appl. Phys. 55, 4015 (1984).

  \bibitem{fukoku} Fukoku-Shinsei Corporation, 2016.

 \bibitem{direct} Aerotech Corporation, www.aerotech.com

\bibitem{snowrsi} D. S. Hussey, D. R. Rich, A. S. Belov, X. Tong, H. Yang, C. Bailey, C. D. Keith, J. Hartfield, G. D. R. Hall, T. C. Black, W. M. Snow, T. R. Gentile, W. C. Chen, G. L. Jones, and E. Wildman, Rev. Sci. Inst. 76, 053503 (2005).
 \bibitem{tong} T. Tong, private communication.
\bibitem{beck} D. Beck, private communication.


U. Schmidt, A. Schnabel, Yu. Sobolev, and K. Tullney, arxiv:1312.3225 (2013).

























  \bibitem{blas-pinningforce}
George S. Park, Charles E. Cunningham, Blas Cabrera, and Martin E. Huber,
Phys.\ Rev.\ Lett. {\bf 68}, 1920 (1992).

\bibitem{simanek-vortexmass}
E.~Simanek,
Phys.\ Lett.\ A, {\bf 194}, 4, p. 323 - 330 (1994);
L.~Janson
American Journal of Physics, {\bf 80}, 2, p. 133-140 (2012).

\bibitem{cryoperm} Amuneal Manufacturing Corp. www.amuneal.com

\bibitem{barnett} S. J. Barnett, Phys. Rev. {\bf 6}, 239 (1915).


\bibitem{patch} C. C. Speake and C. Trenkel, Phys. Rev. Lett. 90, 160403 (2003).




\bibitem{patch2} R. O. Behunin, D. A. R. Dalvit, R. S. Decca, and C. C. Speake, arxiv: 1304.4074 (2013).



\bibitem{MPMS} Magnetic Property Measurement System, Quantum Design, Inc., http://www.qdusa.com

\bibitem{RSIinprep} H. Fosbinder-Elkins, J. Dargert, A.A. Geraci, in preparation

\bibitem{ALPS} Robin Bahre et. al., Any Light Particle Search II Technical Design Report, arxiv:1302.5647 (2013).



\end{thebibliography}
%

\end{document}